\documentclass[prc,amsmath,twocolumn,showkeys,showpacs,superscriptaddress]{revtex4}
\usepackage{graphicx,color}
\usepackage{amssymb}
\usepackage{enumerate}
\usepackage{verbatim}

\begin{document}
  \newcommand {\nc} {\newcommand}
  \nc {\beq} {\begin{eqnarray}}
  \nc {\eeq} {\nonumber \end{eqnarray}}
  \nc {\eeqn}[1] {\label {#1} \end{eqnarray}}
  \nc {\eol} {\nonumber \\}
  \nc {\eoln}[1] {\label {#1} \\}
  \nc {\ve} [1] {\mbox{\boldmath $#1$}}
  \nc {\mrm} [1] {\mathrm{#1}}
  \nc {\half} {\mbox{$\frac{1}{2}$}}
  \nc {\thal} {\mbox{$\frac{3}{2}$}}
  \nc {\fial} {\mbox{$\frac{5}{2}$}}
  \nc {\la} {\mbox{$\langle$}}
  \nc {\ra} {\mbox{$\rangle$}}
  \nc {\etal} {\emph{et al.\ }}
  \nc {\eq} [1] {(\ref{#1})}
  \nc {\Eq} [1] {Eq.~(\ref{#1})}
  \nc {\Ref} [1] {Ref.~\cite{#1}}
  \nc {\Refc} [2] {Refs.~\cite[#1]{#2}}
  \nc {\Sec} [1] {Sec.~\ref{#1}}
  \nc {\chap} [1] {Chapter~\ref{#1}}
  \nc {\anx} [1] {Appendix~\ref{#1}}
  \nc {\tbl} [1] {Table~\ref{#1}}
  \nc {\fig} [1] {Fig.~\ref{#1}}
\newcommand{\figs}[2]{Figs.~\ref{#1} \& \ref{#2}}
  \nc {\bfig} {\begin{figure}}
  \nc {\efig} {\end{figure}}
  \nc {\ex} [1] {$^{#1}$}
  \nc {\Sch} {Schr\"odinger }
  \nc {\flim} [2] {\mathop{\longrightarrow}\limits_{{#1}\rightarrow{#2}}}
  \nc {\IR} [1] {\textcolor{red}{#1}}
  \nc {\IB} [1] {\textcolor{blue}{#1}}
  \nc {\IG} [1] {\textcolor{green}{#1}}

  \nc {\gs} {g.s.}
  \nc {\es} {e.s.}

\title{Asymptotic normalization of mirror states and the effect of couplings}

\author{L.~J.~Titus}
\affiliation{National Superconducting Cyclotron Laboratory, Michigan State University, East Lansing, MI 48824, USA}
\affiliation{Department of Physics and Astronomy, Michigan State University, East Lansing, MI 48824-1321}
\author{P.~Capel}
\affiliation{National Superconducting Cyclotron Laboratory, Michigan State University, East Lansing, MI 48824, USA}
\affiliation{Helmholtz-Insitut Mainz, Johannes Gutenberg-Universit\"at Mainz,
D-55128 Mainz, Germany}
\author{F.~M.~Nunes}
\affiliation{National Superconducting Cyclotron Laboratory, Michigan State University, East Lansing, MI 48824, USA}
\affiliation{Department of Physics and Astronomy, Michigan State University, East Lansing, MI 48824-1321}

\date{\today}

\begin{abstract}
Assuming that the ratio between asymptotic normalization coefficients of mirror states
is model independent, charge symmetry can be used to indirectly extract astrophysically
relevant proton capture reactions on proton-rich nuclei based
on information on stable isotopes. The assumption has been tested for light nuclei within the microscopic cluster model.
In this work we explore the Hamiltonian independence of the ratio between
asymptotic normalization coefficients of mirror states when deformation
and core excitation is introduced in the system. For this purpose we consider a phenomenological $rotor+N$
model where the valence nucleon is subject to a deformed mean field and the core is allowed to excite.
We apply the model to $^8$Li/$^8$B, $^{13}$C/$^{13}$N, $^{17}$O/$^{17}$F,
$^{23}$Ne/$^{23}$Al, and $^{27}$Mg/$^{27}$P.
Our results show that for most studied cases,
the ratio between asymptotic normalization coefficients of mirror states
is independent of the strength and multipolarity of the couplings induced. The exception is for
cases in which there is an $s$-wave coupled to the ground state of the core, the proton system is loosely bound,
and the states have large admixture with other configurations.
We discuss the implications of our results for novae.
\end{abstract}



\maketitle
\section{Introduction}

Novae explosions are a consequence of a thermonuclear runaway on the accreting disk of a white
dwarf within a binary system. The $rp$-process which takes place in {\it novae}, involves reactions
with proton-rich nuclei close to (or at) the proton dripline \cite{review,review2,review3}.
Measuring the corresponding cross sections is particularly challenging,
not only due to the hindrance caused by the Coulomb barrier,
but also due to the fact that they involve rare isotopes (see e.g. \Ref{auria04}).
In many cases, the capture process occurs through specific resonances which
need to be well known \cite{wrede07}. However, even in these cases, it is important to understand
the role of direct capture.

Direct proton captures at low relative energies needed for astrophysics are always peripheral reactions due to
the Coulomb barrier. At the limit of $E \rightarrow 0$ these reactions are uniquely determined by the
asymptotic normalization coefficient (ANC)  of the single proton overlap function of the final nucleus \cite{xu94}.
Based on this realization, the ANC method \cite{xu94} has been put forth as an indirect way of
extracting proton radiative-capture cross sections from ANCs inferred from measurements of nuclear reactions, such as transfer or breakup.

Another indirect technique  \cite{timofeyuk03} uses information on the mirror system.
The idea introduced in \Ref{timofeyuk03} is that charge symmetry can be used to relate the
ANCs of the proton and neutron overlap functions in mirror nuclei.
In this way, while proton capture may
require the knowledge of reactions involving a proton-rich radioactive beam, the neutron counterpart
can be performed with stable beams and thus with much higher accuracy \cite{timofeyuk05a,timofeyuk06}.
In Refs.~\cite{timofeyuk03,timofeyuk05a,timofeyuk06} the ratio ${\cal R}$
of the proton to neutron ANCs squared
is determined for a wide range of light nuclei within a microscopic cluster model (MCM).
This ratio ${\cal R}$ is shown to be independent of the choice for
the NN interaction within a few percent. An analytic derivation of the ratio, ${\cal R}_0$, is also presented \cite{timofeyuk03}.  The ratio obtained from microscopic
calculations is in fair agreement with that predicted by the analytic formula \cite{timofeyuk05a,timofeyuk06}.
Since the original idea was introduced, it has been generalized to resonant states \cite{timofeyuk05b}
and to $\alpha$ cluster states \cite{timofeyuk07}. In this work, we want to explore the
effects of couplings induced by deformation and core excitation in the system.

One might wonder why not calculate the ANC theoretically, instead of relying
on charge symmetry approximations.
The reason for not doing so is the large uncertainty related to the
theoretical prediction of ANCs.
The microscopic calculations presented in Refs.~\cite{timofeyuk03,timofeyuk05a,timofeyuk06} are
strongly dependent on the effective NN interactions used. Ab-initio calculations for light nuclei
are increasingly gaining predictive power, but for the last decade it has been a
true challenge to produce ab-initio overlap functions with a reliable asymptotic behavior for various
technical reasons. The many-body community has put remarkable efforts into extensions of the
traditional methods to enable a good description of the asymptotic behavior. Examples include
i) the coupling of the resonating group method techniques with the no-core shell model (NCSM) \cite{quaglioni08},
ii) expanding the coupled cluster wavefunction in a Breggren basis \cite{hagen10}, and
iii)  using a Green's function method to extract ANCs from Green's function Monte Carlo (GFMC) overlap functions, which have poor asymptotic behavior \cite{nollett11}.
To our knowledge, the work in \Ref{nollett11} consists of the first and only ab-initio ANC calculations
for light nuclei up to A=9, to date.

While ab-initio efforts show promising results, their limitations are hard set:
only light nuclei for NCSM and GFMC and only nuclei around closed shells for
the coupled cluster method. Many nuclei of interest in the $rp$-process are mid-shell nuclei with mass $A>20$
and may have multi-configuration states. It is interesting to explore the effect of couplings
induced by core excitation in such systems.

Effects of including explicitly excited states of the core were studied within
the MCM in Refs.~\cite{timofeyuk05a,timofeyuk06}. It was shown that deviations from the analytic
formula increased. A simple framework of including multi-configuration and excitation in the single nucleon overlap functions
is provided by the $core+N$ phenomenological model \cite{nunes96,esbensen95,vin95}.
In the nineties, this model was applied to a number of light nuclei, including the one-neutron halos
$^{11}$Be \cite{nunes96} and $^{19}$C \cite{ridikas98}.
Starting from a two-body Hamiltonian with an effective deformed $core+N$ interaction which is adjusted to reproduce
the energy levels of the system, one arrives at a coupled-channels equation.
The resulting coupled-channels wavefunction has fragmentation of strength from
the original single particle component to other components involving possible excited states of the core.
Recently, this model was used to explore the connection between the asymptotic properties
of the wavefunction and spectroscopic factors \cite{capel10}. In the present work, we use the model
to study the asymptotic normalization of mirror states and their ratio.

The paper is organized in the following way. In \Sec{theory} we briefly describe the model. Results are presented and discussed in
\Sec{results}, starting with numerical details in \Sec{numerics}, some specific applications to mirror partners in \Sec{specificnuclei}, and
further exploration of the parameter space in \Sec{explore}. Finally in \Sec{conclusions} conclusions are drawn.

\section{Theoretical considerations}
\label{theory}

The $A=B+x$ model introduced in \Ref{nunes96} starts from an effective Hamiltonian representing
the motion of the valence nucleon ($x=n,p$) relative to a core $B$:
\beq
H_A=T_{\ve{r}}+ H_{B}+ V_{Bx}(\ve{r},\xi),
\eeqn{e1}
where $T_{\ve{r}}$ is the relative kinetic energy operator and $H_B$ is the internal Hamiltonian of the core.
The effective interaction between
the core and the valence nucleon depends on the $B$-$x$ relative coordinate $\ve{r}$ but also on
the internal degrees of freedom of the core $\xi$. In this model \cite{nunes96} $V_{Bx}$ is
taken to be a deformed Woods-Saxon potential:
\beq
V_{Bx}(\ve{r})=-V_{ws}\left\{1+\exp\left[\frac{r-R(\theta,\phi)}{a}\right]
\right\}^{-1},
\eeqn{e6}
in which the depth $V_{ws}$ may depend on the $B$-$x$ orbital angular momentum $l$.
Motivated by a deformed shape, the radius $R$ is angle dependent:
\beq
R(\theta,\phi)=R_{ws}[1+\sum_{q=2}^{Q}\beta_q Y_{q0}(\theta,\phi)],
\eeqn{e7}
where $\beta_q$ characterizes the deformation of the core and consequently
the strength of the coupling between various $B+x$ configurations.
As usual, we set $R_{ws}=r_{ws}A^{1/3}$, with $A$ the mass number of the
$B$+$x$ system.
In addition we also include an undeformed spin-orbit coupling term:
\beq
V_{SO}(\ve{r})=\ve{l}\cdot\ve{s}\; V_{so}\frac{1}{r}
\frac{d}{dr}\left[1+\exp\left(\frac{r-R_{ws}}{a}\right)\right]^{-1},
\eeqn{e8}
where $s$ is the spin of the valence nucleon $x$.
When $x=p$, a point-sphere central Coulomb interaction is also included.

The $B+x$ wavefunction is expanded in eigenstates of the core $\Phi_{I^{\pi_B}}$,
with spin $I$, parity $\pi_B$ and eigenenergy $\epsilon_{I^{\pi_B}}$:
\beq
\Psi_{J^\pi}=\sum_{nljI\pi_B}\psi_{nlj}(r){\cal Y}_{lj}(\hat{\ve{r}}) \Phi_{I^{\pi_B}}(\xi).
\eeqn{e2}
Here we factorize the radial part $\psi_{nlj}$
and the spin-angular ${\cal Y}_{lj}$ part for convenience.
The quantum numbers $n$ and $j$ correspond respectively to
the principal quantum number
and the angular momentum obtained from the coupling of the orbital angular
momentum $l$ and the spin $s$. 
Replacing the expansion \eq{e2} into the \Sch equation,
one arrives at a coupled-channel equation \cite{nunes96}:
\begin{equation}
\left[T_r^l+V_{ii}(r)\right]\psi_i(r) +\sum_{j\ne i}V_{ij}(r)\psi_j(r)=
(\varepsilon^x_{J^\pi}-\epsilon_{i})\psi_i(r)
\label{e-cc}
\end{equation}
where $i$ represents all possible $(nljI\pi_B)$ combinations,
$\varepsilon^x_{J^\pi}$ is the relative energy in the $A=B+x$ system
(i.e. same magnitude and opposite sign of the one-neutron or one-proton separation energy),
$T_r^l$ is the radial part of the $B$-$x$ kinetic energy operator,
and the potential matrix elements $V_{ij}$ are
\beq
V_{ij}({r})=\langle\Phi_i(\xi){\cal Y}_i(\hat{\ve{r}})|V_{Bx}(\ve{r},\xi)|
{\cal Y}_j(\hat{\ve{r}})\Phi_j(\xi)\rangle.
\eeqn{e5}
We take $\Phi_{i}$ directly from the rotational model although parameters are fixed
phenomenologically. Solutions of \Eq{e-cc} are found imposing bound-state boundary conditions
and normalizing $\Psi_{J^\pi}$ to unity.
For more details we refer to Refs.~\cite{nunes96,capel10}.

In this model, the norm of $\psi_i$ relates directly to a spectroscopic factor:
\beq
S^x_i=\int_0^\infty |\psi_i|^2r^2dr,
\eeqn{e11}
and the ANC $C^x_i$ is determined from
the asymptotic behavior of $\psi_i$:
\beq
\psi_{i}(r) \flim{r}{\infty} C^x_{i} \,
W_{-\eta^x_i,l+1/2}(2\kappa_i r)
\eeqn{e10}
with $\kappa_i = \sqrt{2\mu_{Bx}|\varepsilon_{J^\pi}-\epsilon_{i}|/\hbar^2}$
and $\mu_{Bx}$ the reduced mass.
The mass of a particle is given by its mass number times $m_N=938.9$~MeV$/c^2$.
In \Eq{e10}, $W$ is the Whittaker function with $\eta^x_i$
the $B$-$x$ Sommerfeld parameter in channel $i$ \cite{AS70}.

To illustrate this model, let us consider the particular mirror pair $^{17}$O/$^{17}$F.
The core of both nuclei is $^{16}$O, which has, apart from the ground state $0^+$,
two low lying states, $2^+$ and $3^-$, coupling strongly to the ground state through
E2 and E3 transitions, respectively.
If one includes in the model space $^{16}$O$(0^+,2^+)$, the ground state of $^{17}$O/$^{17}$F ($5/2^+$) would
not only contain a $1d_{5/2}$ valence nucleon coupled to the ground state $^{16}$O$(0^+)$ but also
for example a $2s_{1/2}$ nucleon coupled to the excited state $^{16}$O$(2^+)$.
A model space containing $^{16}$O$(0^+,3^-)$,
would instead have a $1f_{5/2}$ valence nucleon coupled to the excited state $^{16}$O$(3^-)$,
amongst other orbitals with odd angular momentum.

The main difference between both mirror nuclei is the $B$-$x$ Coulomb interaction.
We should stress that in this work our approach is strongly phenomenological.
Because we are interested in ANCs and these depend strongly on the energy of the system relative
to threshold \cite{sparenberg}, it is essential that we reproduce the experimental separation energies
exactly. Thus, although the initial proton and neutron Hamiltonians only differ by the Coulomb interaction, in our calculations there may be small differences in the adjusted depths
of $V_{Bn}$ and $V_{Bp}$ to reproduce exactly the corresponding binding energies.

As proposed in \Ref{timofeyuk03}, we compare proton ANCs $C_i^p$ with
neutron ANCs $C_i^n$ for mirror states through their ratio
\beq
{\cal R}=\left|\frac{C^p_i}{C^n_i}\right|^2
\eeqn{eR}
In Refs.~\cite{timofeyuk03,timofeyuk07}, a useful analytical approximation of this ratio was derived
\beq
{\cal R}_0=\left| \frac{F_l(i \kappa_i^p R_N)}{\kappa_i^p R_N j_l(i \kappa_i^n R_N)}\right|^2,
\eeqn{e-ratio}
with $F_l$ and $j_l$ being the regular Coulomb function
and the regular Bessel function, respectively \cite{AS70}. The approximation ${\cal R}_0$ is not strongly dependent on the radius of the nuclear interior, $R_N$ \cite{timofeyuk03,timofeyuk05a}.
We will compare our results with the value obtained from this relation.


\section{Results and Discussion}
\label{results}

\subsection{Numerical details}
\label{numerics}

We consider the same cases as in Refs.~\cite{timofeyuk05a,timofeyuk06},
and here present all details concerning the model parameters.
First, it is important to keep in mind that it is not our aim to reproduce all
the properties of these nuclei with our simple model \cite{nunes96},
since in principle microscopic models are much better suited.
Here our aim is to use the $B+x$ model to explore
to what extent core degrees of freedom can modify the picture presented in Refs.~\cite{timofeyuk03,timofeyuk05a,timofeyuk06}.
As the $B$-$x$ interaction is completely phenomenological,
it is essential to have energy levels to constrain the interaction.
Below we provide details of the fitting for each case.
Core excitation energies are taken from the database of the
National Nuclear Data Center \cite{NNDC}.
It is the deformation that introduces tensor components in the
interaction and that allows for configuration admixture between various core states.
Values for the deformation parameters for each case, as well as the
states to be considered in the coupled channel equation, are given.
The geometry for the Woods-Saxon interaction and the strength of
the spin-orbit force $V_{so}$ are fixed at constant values
(see Secs.~\ref{specificnuclei} and \ref{explore}).
The depth of the central potential $V_{ws}$ is then adjusted to
reproduce the $B$-$x$ separation energy (shown in Table~\ref{depths}).
In some cases we fit more than one state per nucleus.
This introduces an $l$-dependence in $V_{ws}$.
All calculations are performed with the program {\sc face} \cite{face}.

\paragraph{\rm $^8$Li/$^8$B:}
The $B+x$ description of these mirror nuclei corresponds to
$^7$Li$+n$ for $^8$Li and $^7$Be$+p$ for $^8$B. The respective $B$-$x$ relative
energies are $\varepsilon^n_{2^+}=-2.032$~MeV
and $\varepsilon^p_{2^+}=-0.1375$ MeV.
The $2^+$ ground state of these nuclei is described as a dominant
$1p_{3/2}$ nucleon bound to the $3/2^-$ ground state of the core.
The $1/2^-$ state of the core is also considered
($\epsilon_{1/2^-}(^7{\rm Li})=0.478$~MeV and
$\epsilon_{1/2^-}(^7{\rm Be})=0.429$~MeV).
The quadrupole deformation that couples both core states
of $^7$Li is $\beta_2=0.34$.
That of $^7$Be, being predicted to be around 0.3--0.4 \cite{pudliner97},
is chosen equal to that of $^7$Li.

\paragraph{\rm $^{13}$C/$^{13}$N:}
In this mirror pair the core is $^{12}$C for both nuclei.
The dominant configuration of the $1/2^-$ ground state is a
$1p_{1/2}$ nucleon coupled to the $0^+$ ground state of $^{12}$C.
The relative $^{12}$C-$x$ energies are $\varepsilon^n_{1/2^-}=-4.946$ MeV and $\varepsilon^p_{1/2^-}=-1.944$ MeV for $^{13}$C and $^{13}$N respectively.
The $2^+$ excited state of $^{12}$C at $\epsilon_{2^+}=4.439$~MeV
is also considered with the coupling $\beta_2=-0.6$ \cite{vermeer83}.

\paragraph{\rm $^{17}$O/$^{17}$F:}
For this $^{16}$O$+x$ mirror pair, our model reproduces both the
$5/2^+$ ($\varepsilon^n_{5/2^+}=-4.144$~MeV or $\varepsilon^p_{5/2^+}=-0.601$~MeV)
and $1/2^+$ ($\varepsilon^n_{1/2^+}=-3.273$~MeV or $\varepsilon^p_{1/2^+}=-0.106$~MeV) bound states as predominantly
$1d_{5/2}$ and $2s_{1/2}$ valence nucleons coupled to the $0^+$ ground state
of $^{16}$O.
For $^{16}$O we consider the effect of the coupling between
the $0^+$ ground state and either
the $2^+$ excited state at $\epsilon_{2^+}=6.917$~MeV or
the $3^-$ excited state at $\epsilon_{3^-}=6.129$~MeV.
The corresponding quadrupole and octopole deformations are
$\beta_2=0.36$ \cite{raman01} and $\beta_3=0.75$ \cite{spear89}, respectively.
To adjust both $5/2^+$ and $1/2^+$ states, $V_{ws}$ in the
$s$ and $d$ wave differ slightly.
When considering the coupling to the $3^-$ excited state of \ex{16}O,
we set the depths of the potential in the negative-parity partial waves
according to $V_{ws}(l=1)=V_{ws}(l=2)$ and $V_{ws}(l=3)=V_{ws}(l=0)$.
In this way, the partial waves corresponding to the dominant configurations
in the $5/2^+$ and $1/2^+$ states have the same potential depth.

\begin{table*}
\caption{Depths $V_{ws}$ of the central potential for the various cases
listed in \Sec{numerics} (values are given in MeV). The first number is
the depth for the neutron case, and the second number is for the proton case.}\label{depths}
\begin{tabular}{c|c|c|c||c|c}\hline\hline
& & \multicolumn{2}{|c||}{$r_{ws}=1.2$~fm $a=0.5$~fm} & \multicolumn{2}{c}{$r_{ws}=1.25$~fm $a=0.65$~fm} \\ 
 & $I^{\pi_B}$ & $V_{so}=6$~MeV & $V_{so}=8$~MeV & $V_{so}=6$~MeV & $V_{so}=8$~MeV\\ \hline
 
$^8$Li/$^8$B& $3/2^-,1/2^-$ & & & & \\ 
$V_{ws}$ & & $58.9168$/$59.6080$ & $61.5840$/$62.3011$ & $42.6722$/$42.7479$ & $41.5371$/$42.0051$\\\hline

$^{13}$C/$^{13}$N& $0^+,2^+$ & & & & \\ 
$V_{ws}$ & & $59.6504$/$60.4455$ & $59.6635$/$60.4436$ & $56.4058$/$56.8085$ & $56.3261$/$56.7085$\\\hline

$^{17}$O/$^{17}$F & $0^+,3^-$ & & & & \\ 
$V_{ws}$ ($l=0$ and 3) & & $51.6320$/$51.8248$ & $51.5582$/$51.7569$ & $47.6517$/$47.2511$ & $47.6102$/$47.2151$\\
$V_{ws}$ ($l=1$ and 2) & & $60.3527$/$61.1595$ & $59.5078$/$60.3983$ & $57.0489$/$57.3955$ & $56.4297$/$56.8736$\\\hline

$^{17}$O/$^{17}$F & $0^+,2^+$ & & & & \\ 
$V_{ws}$ ($l=0$ and 3) & & $52.3626$/$53.1800$ & $52.9035$/$53.6801$ & $48.5662$/$48.6508$ & $48.9239$/$48.9721$\\
$V_{ws}$ ($l=1$ and 2) & & $53.8238$/$54.2036$ & $52.1111$/$52.4642$ & $51.2895$/$51.2182$ & $49.8253$/$49.7276$\\\hline

$^{23}$Ne/$^{23}$Al& $0^+,2^+,4^+$ & & & & \\ 
$V_{ws}$ ($l=0$ and 3) & & $54.4839$/$54.4839$ & $54.6976$/$54.6976$ & $49.0477$/$49.0477$ & $49.3356$/$49.3356$\\
$V_{ws}$ ($l=1$ and 2) & & $55.4659$/$56.2811$ & $53.9916$/$54.7628$ & $52.6434$/$52.7345$ & $51.3534$/$51.4028$\\\hline

$^{27}$Mg/$^{27}$P& $0^+,2^+,4^+$ & & & & \\ 
$V_{ws}$ ($l=0$ and 3) & & $52.5428$/$53.2944$ & $51.2411$/$51.9676$ & $48.9469$/$48.2467$ & $47.7296$/$46.9657$\\
$V_{ws}$ ($l=1$ and 2) & & $56.4027$/$56.4027$ & $56.9672$/$56.9672$ & $53.6536$/$53.6536$ & $54.1691$/$54.1691$\\
\hline\hline

\end{tabular}
\end{table*}

\paragraph{\rm $^{23}$Ne/$^{23}$Al:}
The cores in this mirror pair are $^{22}$Ne and $^{22}$Mg.
Our model reproduces the $5/2^+$ ground state of both nuclei
with $\varepsilon^n_{5/2^+}=-5.200$~MeV or $\varepsilon^p_{5/2^+}=-0.122$~MeV,
and the $1/2^+$ excited state of $^{23}$Ne with $\varepsilon^n_{1/2^+}=-4.184$ MeV.
The configuration of the ground state is dominated by a $1d_{5/2}$ nucleon
bound to the $0^+$ ground state of the core. The excited state of $^{23}$Ne is
mostly a $2s_{1/2}$ neutron bound to $^{22}$Ne$(0^+)$.
We consider couplings between the lowest $0^+$, $2^+$ and $4^+$ core states,
with excitation energies $\epsilon_{2^+}=1.274$~MeV and $\epsilon_{4^+}=3.357$~MeV for $^{22}$Ne and
$\epsilon_{2^+}=1.247$~MeV and $\epsilon_{4^+}=3.308$~MeV for $^{22}$Mg.
These three states are described as the first three levels of one rotational
band with deformation parameters $\beta_2=0.58$ \cite{raman01} and $\beta_2=0.562$ \cite{raman01}
for $^{22}$Mg and $^{22}$Ne, respectively.
To reproduce the two energy levels in $^{23}$Ne, we need to consider a slight difference
between $V_{ws}(l=0)$ and $V_{ws}(l=2)$.
The same value for $V_{ws}(l=0)$ is used in $^{23}$Al with small adjustments
made to $V_{ws}(l=2)$ to reproduce the binding energy exactly.

\paragraph{\rm $^{27}$Mg/$^{27}$P:}
In these $^{26}$Mg$+n$ and $^{26}$Si$+p$ mirror systems, we
reproduce the $1/2^+$ ground states as a dominant $2s_{1/2}$ nucleon
bound to the $0^+$ ground state of the core by $\varepsilon^n_{1/2^+}=-6.443$~MeV or $\varepsilon^p_{1/2^+}=-0.861$~MeV.
For the neutron system, we also consider the excited state $3/2^+$
with $\varepsilon^n_{3/2^+}=-5.459$~MeV to pin down the $d$-wave potential
as its configuration is dominated by a $1d_{3/2}$ neutron
bound to $^{26}$Mg($0^+$). Here, we consider couplings between the first $0^+$, $2^+$ and $4^+$ core states,
with excitation energy $\epsilon_{2^+}=1.8$~MeV for both cores and $\epsilon_{4^+}=4.32$~MeV for $^{26}$Mg
and $\epsilon_{4^+}=4.18$~MeV for $^{26}$Si.
Deformation parameters are $\beta_2=0.482$ \cite{raman01} and $\beta_2=0.446$ \cite{raman01}
for $^{26}$Mg and $^{26}$Si, respectively.
To reproduce the energy levels in $^{27}$Mg, different depths $V_{ws}$ are taken
for $l=0$ and $l=2$. For $^{27}$P,  the same $V_{ws}(l=2)$ is used as for $^{27}$Mg
but small adjustments are made to $V_{ws}(l=0)$ to obtain the correct binding energy.

\subsection{Ratio for specific mirror partners}
\label{specificnuclei}

For comparison with previous works, we fix the deformation of the core, adjust the depth $V_{ws}$ of the interaction
to reproduce binding energies as detailed in Section \ref{numerics}, and solve the coupled channels equation.
To evaluate the sensitivity of our calculations to the choice of the
$B$-$x$ potential,
we consider two geometries for the mean field, namely radius $r_{ws}=1.2$ fm and diffuseness $a=0.5$ fm and
radius $r_{ws}=1.25$ fm and diffuseness $a=0.65$ fm. We first fix $V_{so}=6$ MeV with the same geometry as
the Woods-Saxon potential, but repeat the calculations for the choice of $V_{so}=8$ MeV.
The depths $V_{ws}$ obtained for each of the cases listed in \Sec{numerics}
are given in Table~\ref{depths}.

From the resulting proton and neutron wavefunctions, we determine ANCs and the ratio ${\cal R}$ \eq{eR}.
The ratio ${\cal R}$ for the dominant component for each case is shown in Table \ref{ratios} and
corresponds to $r_{ws}=1.25$ fm, $a=0.65$ fm and $V_{so}=6$ MeV.  The uncertainty reflects the range obtained with the other geometry and spin-orbit strength.
Our values for $\cal R$ are compared to the
values obtained from the analytic formula ${\cal R}_0$ \eq{e-ratio} (using the experimental binding energies
and $R_N=1.25 A^{1/3}$) and those obtained within the MCM,
assuming two clusters and taking the Minnesota interaction ${\cal R}_{MCM}$ \cite{timofeyuk05a,timofeyuk06}.

\begin{table*}[t]
\caption{Ratio of proton to neutron ANCs for the dominant component: comparison of this work ${\cal R}$ with
the results of the analytic formula ${\cal R}_0$ \eq{e-ratio} and the results of
the microscopic two-cluster calculations  ${\cal R}_{MCM}$ \cite{timofeyuk05a,timofeyuk06} including the Minnesota interaction.
The uncertainty in ${\cal R}$ account for the sensitivity to the parameters of $V_{Bx}$.
}
\label{ratios}
\begin{center}
\begin{tabular}{|c|c c c c c|}
\hline
nuclei                  & $I^{\pi_B}$ 		 & $nlj$	& ${\cal R}$       & ${\cal R}_0$&  ${\cal R}_{MCM}$\\
\hline
$^{8}$Li/$^{8}$B  	& $3/2^-,1/2^-$     & $1p3/2$	& $1.04 \pm 0.04$  & 1.12 & $1.08$	\\
$^{13}$C/$^{13}$N 	& $0^+,2^+$         & $1p1/2$	& $1.19 \pm 0.02$  & 1.20 & $1.14$ \\
$^{17}$O/$^{17}$F (g.s.)& $0^+,3^-$ 	& $1d5/2$	& $1.18 \pm 0.01$  & 1.22 & $1.19$ \\
$^{17}$O/$^{17}$F (e.s.)& $0^+,3^-$ 	& $2s1/2$	& $693  \pm 16$    & 799  & $736$  \\
$^{17}$O/$^{17}$F (g.s.)& $0^+,2^+$ 	& $1d5/2$	& $1.219 \pm 0.004$& 1.22 & $1.19$ \\
$^{17}$O/$^{17}$F (e.s.)& $0^+,2^+$ 	& $2s1/2$	& $756  \pm 23$    & 799  & $736$  \\
$^{23}$Ne/$^{23}$Al     & $0^+,2^+,4^+$   & $1d5/2$	
				& \;\; $(1.852  \pm 0.014)\times10^4$ \;   & \; $2.06\times10^4$ \; & \; $2.96\times10^4$  \\
$^{27}$Mg/$^{27}$P      & $0^+,2^+, 4^+$ 	& $2s1/2$	& $40.1  \pm 1.8$   & 43.7 & $44.3$ \\
\hline
\end{tabular}
\end{center}
\end{table*}

For the first three cases studied, namely $^8$Li/$^8$B, $^{13}$C/$^{13}$N, and
$^{17}$O/$^{17}$F(g.s.) our ratios are very close to the values obtained with the analytical
formula and those obtained within the MCM. Larger deviations are found for $^{17}$O/$^{17}$F(e.s),
$^{23}$Ne/$^{23}$Al and $^{27}$Mg/$^{27}$P. While in $^{17}$O/$^{17}$F(e.s) the core in the neutron and
proton systems are the same, in the last two cases the core $\beta_2$ differs slightly.
The deviations with the analytic formula and MCM are not caused by this difference. 

For $^{23}$Ne/$^{23}$Al, it is important to note
that in our calculations we impose realistic binding
energies whereas in the MCM results, binding energies can sometimes differ significantly.
Since ${\cal R}$ depends strongly on the binding energies, this can be the cause for the large difference
between our values and those of \Ref{timofeyuk06}). The values of  ${\cal R}_0$ presented
in Table~\ref{ratios} also assume the experimental binding energies and therefore differences between
${\cal R}$ and ${\cal R}_0$ must be related to the failure of the simple analytical relation.

One could presume that the examples for which our model predicts significantly different
ratio than the analytic prediction and the MCM are those in which the admixture
with core excited configurations are largest. This is not the case: large admixture, or small
spectroscopic factors, alone are not sufficient to cause a deviation from ${\cal R}_0$ or
previously calculated ${\cal R}_{MCM}$. Spectroscopic factors are around: 0.9 for $^8$Li/$^8$B,
0.3 for $^{13}$C/$^{13}$N, 0.6-0.9 for $^{17}$O/$^{17}$F(g.s.), 0.7-0.9 for $^{17}$O/$^{17}$F(e.s.),
0.7 for $^{23}$Ne/$^{23}$Al and 0.5 for $^{27}$Mg/$^{27}$P.
What can be remarked is that the largest discrepancies appear for the cases in which the proton
is very loosely bound. Another remarkable point is that our predicted ratio is always smaller
than the analytical estimate. This feature is further investigated in the following section. \\

\subsection{Exploring the parameter space}
\label{explore}

In this subsection, we use the deformation parameter as a free variable to explore different
physical situations beyond the particular nuclei used as test cases. The configurations of
the $^{23}$Ne/$^{23}$Al and $^{27}$Mg/$^{27}$P pairs being very similar to those
of the $^{17}$O/$^{17}$F systems in its ground state and excited state, respectively, 
we concentrate on the three lighter cases.
Given the range of values for the deformation parameters, we vary the deformation
between 0 and 0.7. For each deformation parameter, energies for the proton and neutron
systems were refitted by small adjustments of $V_{ws}$ to eliminate erroneous variations
of the ANC due to changes in the binding energies: overall $V_{ws}^p \approx V_{ws}^n$.
We fix the geometry: the standard $r_{ws}=1.25$ fm and $a=0.65$ fm for $^8$Li/$^8$B \cite{schumann06},
$r_{ws}=1.14$ fm and $a=0.5$ fm for $^{13}$C/$^{13}$N \cite{nunes96}, and
$r_{ws}=1.2$ fm and $a=0.64$ fm for $^{17}$O/$^{17}$F \cite{sparenberg2000}.
The geometry for the spin-orbit force is taken to be the same as for the
nuclear force, and the depth is fixed at around $6$ MeV, for all cases.

We find no significant difference in the
ratio ${\cal R}$ for both $^{8}$Li/$^{8}$B and $^{13}$C/$^{13}$N mirror pairs. In these
cases the main components of the wavefunction are $p$ waves,
even in the configurations including core excitation.
For $\left|\beta_2\right|=0.0$--0.7
the resulting range of values for ${\cal R}$ are: (1.038--1.044) for $^{8}$Li/$^{8}$B
and (1.201--1.251) for $^{13}$C/$^{13}$N.
This constancy is obtained even though the variation in $\beta$ leads to significant changes in the spectroscopic
factor: $S^x_{1p_{3/2}}$ goes from 1 to 0.75 for $^{8}$Li/$^{8}$B, while
$S^x_{1p_{1/2}}$ decreases down to 0.32 for $^{13}$C/$^{13}$N.
Even if the system is made artificially less bound,
the variation of ${\cal R}$ remains small and within
the uncertainties of the geometry parameters for the interaction.
The significant stability of ${\cal R}$ with such large changes in both deformation
and admixture of different configurations suggests a universality of the
mirror technique developed in \Ref{timofeyuk03}.

\begin{figure}[t!]
{\centering \resizebox*{0.4\textwidth}{!}{\includegraphics{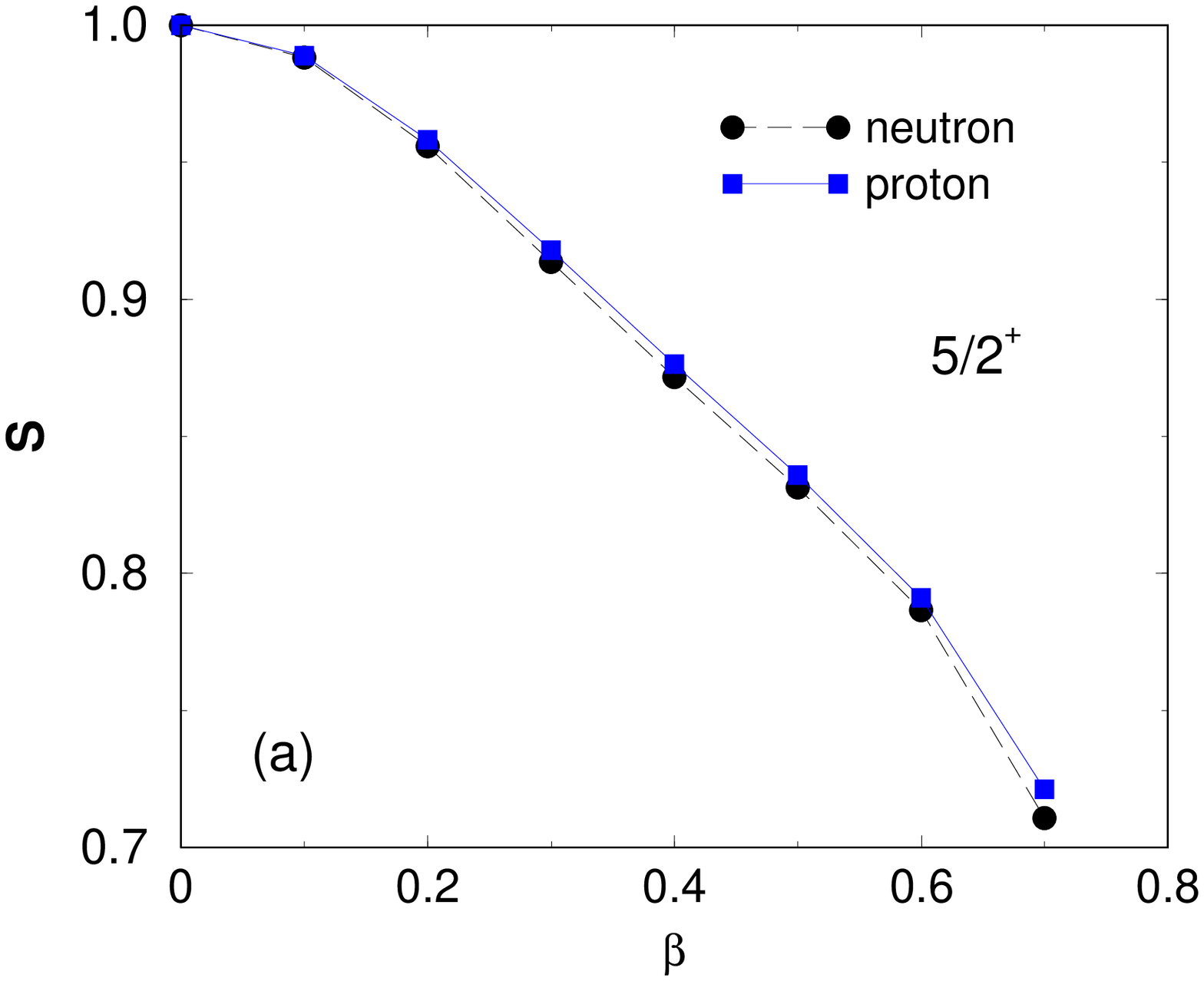}}} \\
{\centering \resizebox*{0.4\textwidth}{!}{\includegraphics{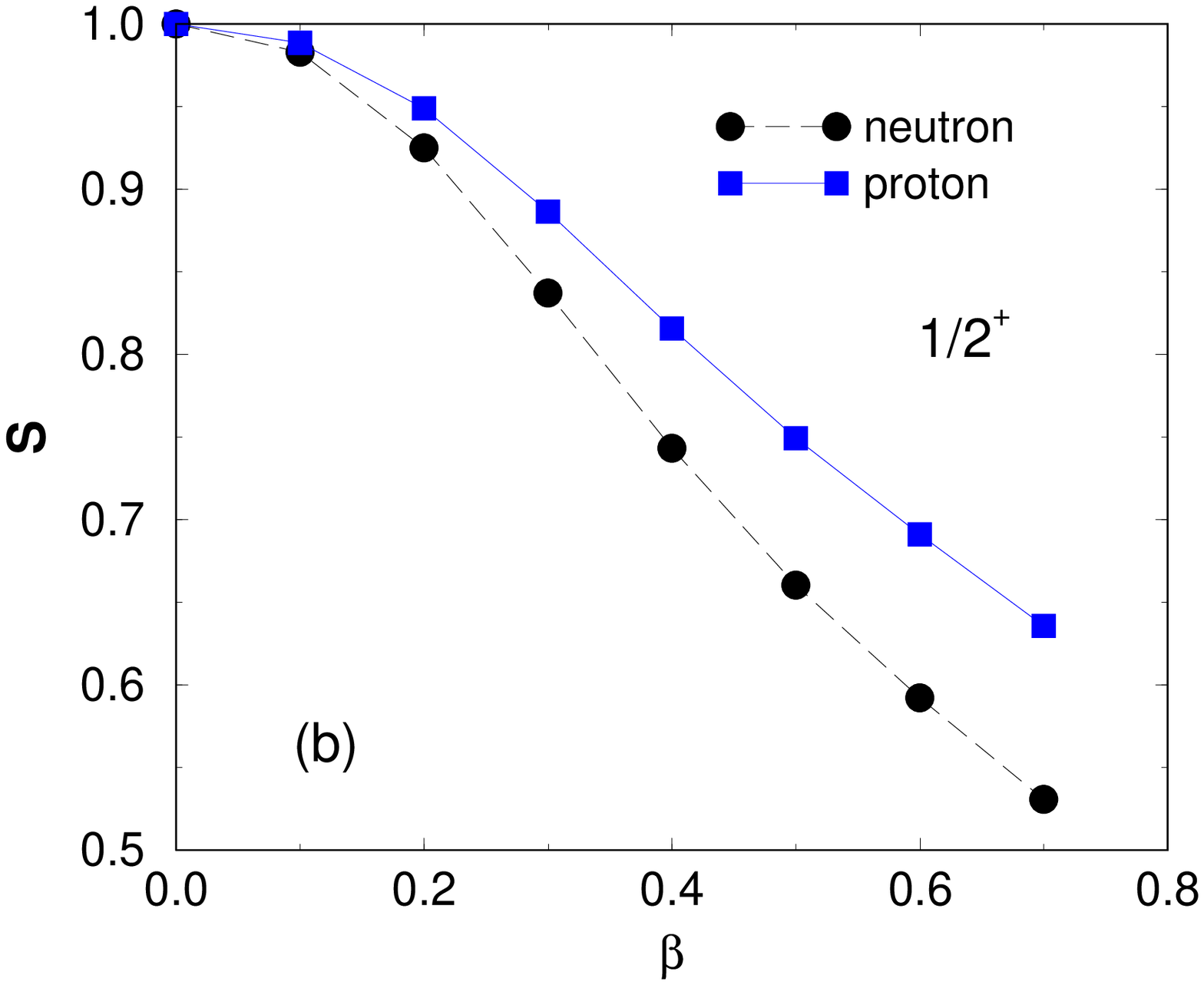}}} \\
\caption{\label{sf} (Color online) Neutron and proton spectroscopic factors
for $^{17}$O and $^{17}$F, respectively, considering the \ex{16}O core in its
$0^+$ ground state and $2^+$ first excited state:
(a) $5/2^+$ ground state and
(b) $1/2^+$ first excited state.}
\end{figure}

The situation for $^{17}$O/$^{17}$F is different. In this case core excitation introduces different orbital
angular momenta in the wavefunction. We consider the separate effect of including the $3^-$ state
and the $2^+$ state. Let us first consider the inclusion of $^{16}$O($0^+, 3^-$).
For each $\beta_3$, energies for the two lowest states in $^{17}$O and $^{17}$F were refitted by small adjustments
of $V_{ws}(l=0)$ and $V_{ws}(l=2)$.
As mentioned in \Sec{numerics}, the depth of the potential in the negative-parity
partial waves is set to $V_{ws}(l=1)=V_{ws}(l=2)$, and $V_{ws}(l=3)=V_{ws}(l=0)$.
In this way, all the depths were constrained phenomenologically.
Here again the variations in ${\cal R}$ are small.
Even though over 30\% of the $5/2^+$ ground-state wave function is in a
core-excited configuration at $\beta_3=0.7$,
the change in ${\cal R}$ is less than 2\%.
For this $\beta_3$, the $1/2^+$ excited-state wave function is almost exclusively in the
$^{16}$O$(0^+)\otimes 2s_{1/2}$ configuration ($S^x_{2s_{1/2}}\approx 95$\%).
Expectedly, the change in the corresponding ratio is limited to less than 1\%.

\begin{figure}[t!]
{\centering \resizebox*{0.4\textwidth}{!}{\includegraphics{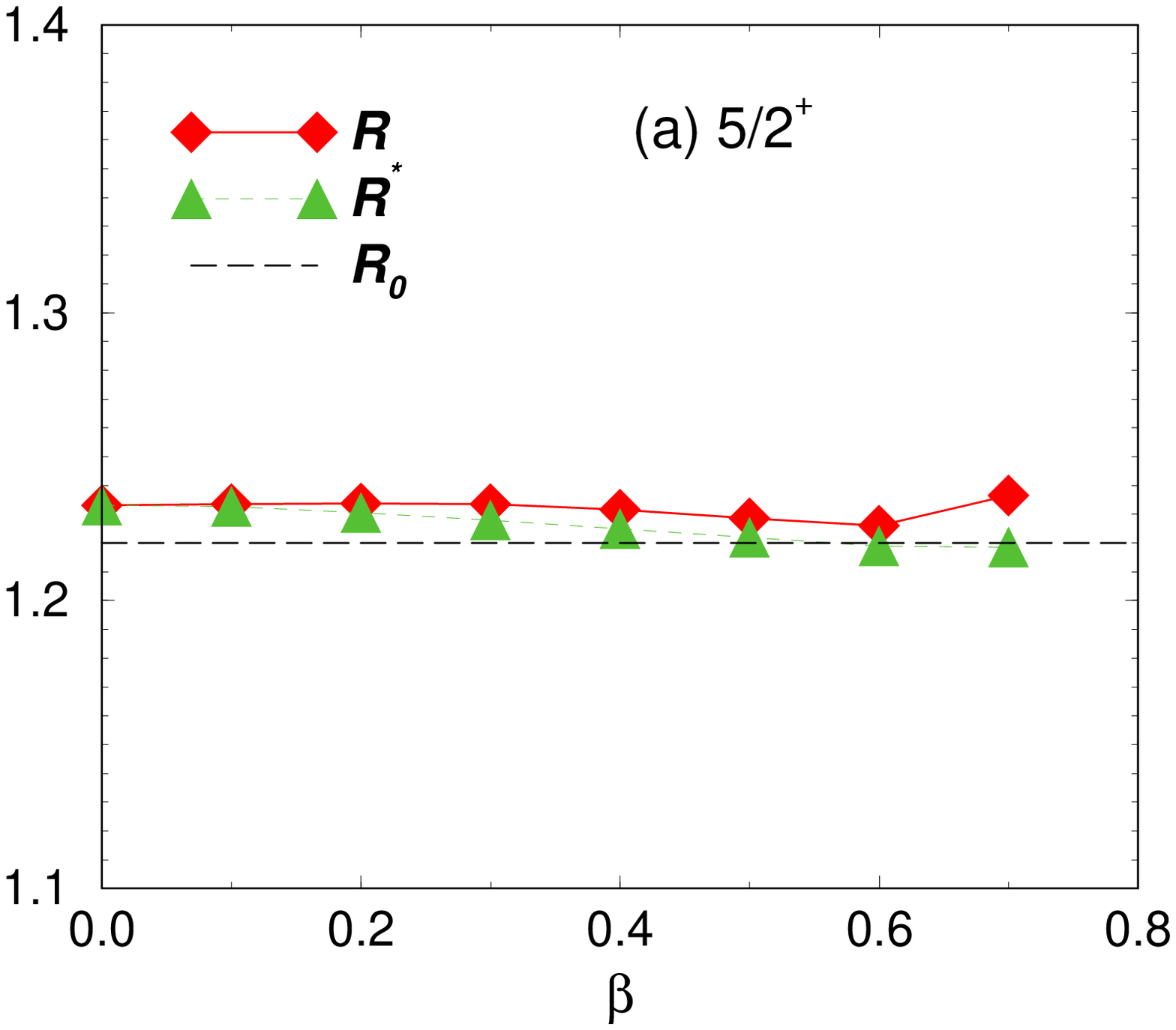}}} \\
{\centering \resizebox*{0.4\textwidth}{!}{\includegraphics{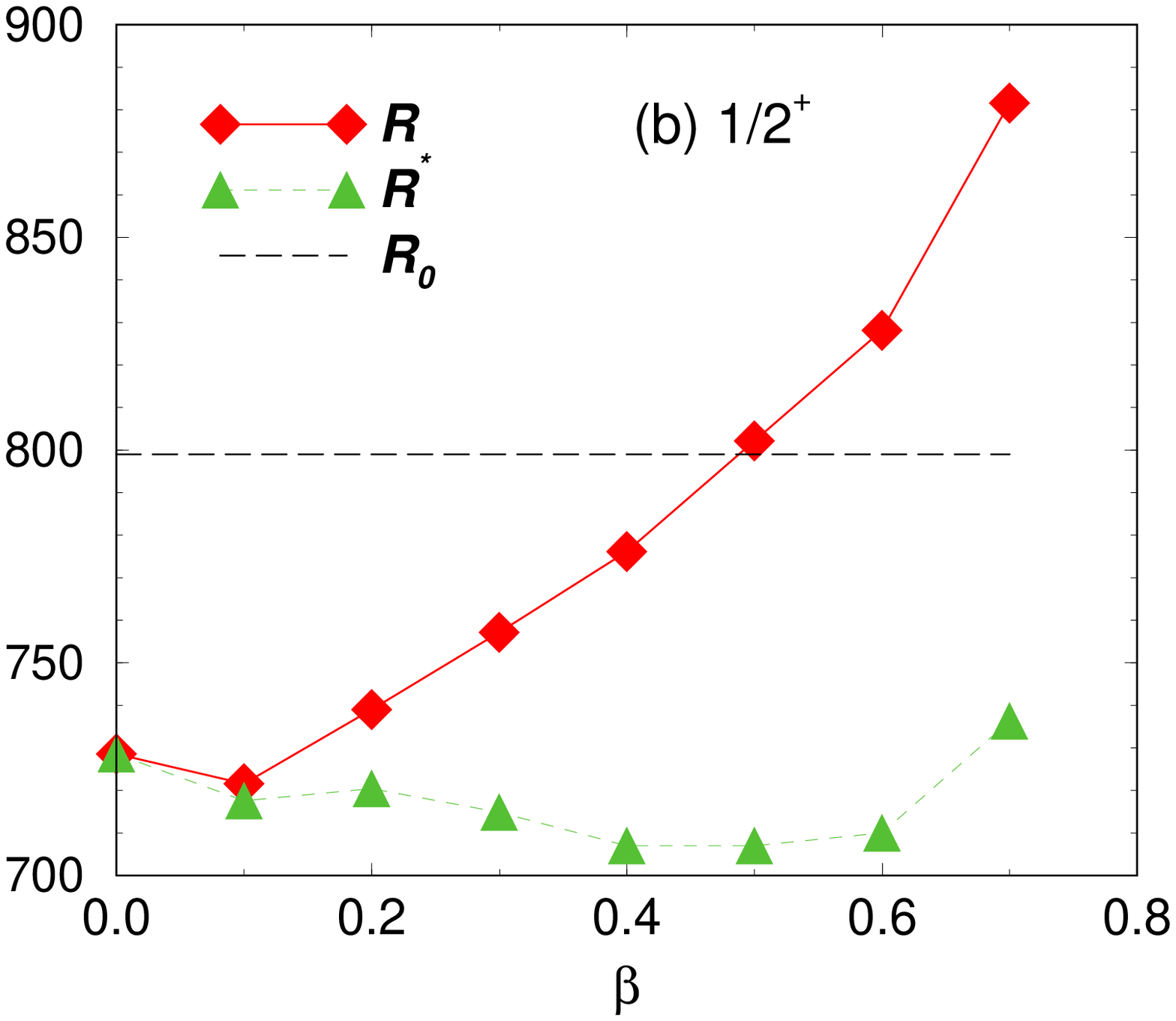}}} \\
\caption{\label{ratio} (Color online) Ratio of proton and neutron ANCs
for $^{17}$O and $^{17}$F, respectively, including \ex{16}O($0^+, 2^+$):
(a) $5/2^+$ ground state and
(b) $1/2^+$ first excited state. }
\end{figure}
Next we consider the inclusion of $^{16}$O($0^+, 2^+$). In this case the $d_{5/2}$ ground state admixes
with an $s_{1/2}$ component with the core in its excited state, while in the $1/2^+$ state, the $s_{1/2}$ coupled
to the g.s. core admixes with $d$ components with the core in its $2^+$ state.
Again, energies for the two lowest states in $^{17}$O and $^{17}$F were
refitted by simultaneously adjusting $V_{ws}(l=0)$ and $V_{ws}(l=2)$ for each $\beta_2$.
For both $5/2^+$ and $1/2^+$  states,
the spectroscopic factor \eq{e11} of the dominant component (which has the core in its ground state) suffers a large reduction at large $\beta_2$,
as shown in Fig.~\ref{sf}. While for the ground state, the proton and neutron spectroscopic factors vary together (Fig.~\ref{sf}a),
for the excited state it becomes clear that the admixture in the neutron system is larger than in the proton system (Fig.~\ref{sf}b).
This is then reflected in a different behavior of the ANC ratios.
In \fig{ratio} we present the ratio $\cal R$ \eq{eR}, as well as a
modified ratio compensating for the changes in spectroscopic factors ${\cal R}^*={\cal R} S^n/S^p$.
The analytical prediction ${\cal R}_0$ \eq{e-ratio} is also shown (horizontal dashed lines).
For the $5/2^+$ ground state, neither $\cal R$ nor $\cal R^*$
deviate much from the value at $\beta_2=0$,
corresponding to the single particle prediction (Fig.~\ref{ratio}a).
They are also very close to the analytical prediction, ${\cal R}_0$.
On the contrary,
for the $1/2^+$ excited state, ${\cal R}$ shows a large variation, mainly, but not only, caused by the difference between neutron and proton
spectroscopic factors (Fig.~\ref{ratio}b), as expected from the results of \Ref{capel10}. This can be deduced from their relative variations across the considered $\beta_2$ range: while ${\cal R}$ varies by $22\%$,
${\cal R}^*$ varies by less than $3\%$.
They also differ more from ${\cal R}_0$.
As noted in the MCM studies \cite{timofeyuk05a,timofeyuk06},
the ratio $\cal R$ at the realistic deformation of the $^{16}$O core
(i.e. $\beta_2=0.36$)  is well approximated by the average between
${\cal R}_0$ and the single-particle ratio, i.e. ${\cal R}$ at $\beta_2=0$.
Since this result is strongly dependent on the value of the deformation,
we do not believe it can be safely generalized to other systems.
The features illustrated in \fig{ratio} can be directly extrapolated to $^{23}$Al and $^{27}$P.
As mentioned before, the former has a structure very similar to that of $^{17}$F(g.s.),
while the latter exhibits the same components as $^{17}$F(e.s.).

In Refs.~\cite{timofeyuk05a,timofeyuk06} core excitation is explored within the MCM. Already then there was
growing disagreement between ${\cal R}_{MCM}$ and ${\cal R}_0$ as more core states were explicitly included in the model space.
This was understood in terms of the long range Coulomb quadrupole term
added to the Hamiltonian in the proton case, a term not considered in the derivation of ${\cal R}_0$,
nor in our present calculations.
Here however, we not only see a deviation from ${\cal R}_0$, but also a strong dependence on the deformation parameter for particular cases. Therefore we conclude the source for deviations
from ${\cal R}_0$ and the break down of the constant ratio concept is induced by the nuclear
quadrupole term, which is present in both neutron and proton systems.

The surprising results for the $1/2^+$ mirror states led to several additional tests which isolated the cause for the large coupling dependence in ${\cal R}$.
There are three essential ingredients:
low binding, the existence of an $s$-wave component coupled to the ground state of the core, and a significant admixture with other configurations. It appears that when all three conditions are met, the differences between the neutron and proton wavefunctions increase around the surface, exactly where the
nuclear quadrupole interaction peaks. This results in a stronger effect of coupling on the neutron
system compared to the proton system, inducing differences in $S^n$ relative to $S^p$, which
reflect on a coupling dependence in $\cal R$. Our tests show that the effect is independent on whether the
wavefunctions have a node.

\section{Conclusions}
\label{conclusions}

A proposed indirect method for extracting proton capture rates from neutron mirror partners
relies on the ratio between asymptotic normalization coefficients of the mirror states
being model independent. In this work, we test this idea against core deformation
and excitation. We consider a  $core+N$ model where the core is
deformed and allowed to excite and apply it to a variety of mirror pairs ($^8$Li/$^8$B, $^{13}$C/$^{13}$N, $^{17}$O/$^{17}$F, $^{23}$Ne/$^{23}$Al, and $^{27}$Mg/$^{27}$P.). 
We stress that our approach is strongly phenomenological: for each case we always fit the
neutron and proton binding energies exactly. This is not the approach followed in previous works \cite{timofeyuk05a,timofeyuk06,timofeyuk05b}. Imposing instead equal nuclear interactions $V_n=V_p$ in our model would lead to a strong and erroneous
deformation dependence of ${\cal R}$ due to unequal changes in the neutron and proton binding energies.
In that case, even ${\cal R}_0$ would become model dependent.

We explored how the mirror states evolve as a function of deformation (coupling strength).
For most cases the ratio of the ANC of mirror states was found to be independent
of the deformation. From our investigations we conclude that there are three conditions that need to be met for the idea of a model-independent ratio to break down with deformation or core excitation:
i) the proton system should have very low binding, ii) the main configuration should be an $s$-wave component coupled to the ground state of the core, and iii) there should be significant admixture with other configurations.
This has implications for the application of the  indirect method based on the ANC ratio
to reactions relevant to novae, namely pertaining the direct capture
component of  $^{26}$Si(p,$\gamma$)$^{27}$P. In connecting the ANC of $^{27}$Mg
and $^{27}$P one should be careful with coupling between different configurations.

An analytic formula for the ratio ${\cal R}_0$ was derived \cite{timofeyuk03} using
a single particle configuration for neutron and proton states.
In \cite{timofeyuk05a} it is suggested that differences between ${\cal R}_0$ and ${\cal R}$
calculated within MCM arose due to  the quadrupole Coulomb interaction, which is not included
in the proton state, when deriving ${\cal R}_0$, but of course is included in the MCM calculations.  
We do not include this term in our calculations and yet still find deviations between
our ${\cal R}$ and ${\cal R}_0$. These can only be due to the nuclear
quadrupole term.

When an incoming $s$-wave neutron is involved one should choose an adequate probe to measure it.
While $s$-wave proton capture (usually to a bound $p$-state) is a peripheral process for the low relative energies of astrophysical interest, the $s$-wave neutron capture is not and generally depends on the whole overlap function. Nevertheless, in principle one can extract ANCs for the neutron system from peripheral nuclear reactions (transfer or breakup) using
an appropriate choice of kinematic conditions. That ANC would then relate to the astrophysically relevant proton ANC.

We are grateful to Natasha Timofeyuk for suggesting this project and providing important
feedback on the work and we thank Ron Johnson for many useful discussions.
This work was partially supported by the National Science Foundation
grant PHY-0555893, the Department of Energy through grant DE-FG52-08NA28552
and the TORUS collaboration DE-SC0004087.


\end{document}